\documentclass [a4paper, 12pt]{article}
\usepackage{amssymb}
\usepackage[dvips]{graphicx}
\title{Measurements of the optical mirror coating properties.}

\author{V.B. Braginsky, A.A. Samoilenko \\ Dept. of Phys. Moscow
State University\\ Moscow 119992, Russia.}

\begin{document}
\maketitle

\begin{abstract}
The results of measurement of optical mirror coating are
presented. These results indicate that Standard Quantum Limit
of sensitivity can be reached in the second stage of LIGO
project if it is limited by thermoelastic noise in the coating
only.
\end{abstract}

\section{Introduction.}
The sensitivity of terrestrial laser interferometer gravitational
wave antennae \cite{Abram} crucially depends on how the analysis
of different kind of noises in parts of the set up, especially in
the optical mirrors (test masses), is in-depth. Few years ago the
analysis of the mirrors thermoelastic and thermorefractive noises
''contribution'' to the total noise budget was carried out
\cite{B1, Th, B2}. The net result of this analysis is: these types
of noises can make a significant ''contribution'' to the total
noise if wrong material will be chosen for the mirrors. The
analysis has also shown that fused silica has a serious advantage
because of very small thermal expansion coefficient $\alpha_{\rm
Si\, O_2} \simeq5\times10^{-7} \rm\: K^{-1}$. At the same time the
role of the mirrors coating has not been taken into account in
this analysis. The coating of the mirror provides very important
parameter of the mirror: the difference between the reflectivity
$R$ and unity. The smaller is the value of $(1-R)$, the higher is
the sensitivity of the antenna. It is worth noting that the
smallness of $(1-R)$ also defines the sensitivity in many high
resolution spectroscopical measurements.

The analysis of the role of thermoelastic noise produced due to
the coating has been carried out recently by S.P.~Vyatchanin and
by one of this article authors and independently by M.~Fejer and
S.~Rowan \cite{BV, Row}. The analytical formulas for the values of
spectral densities of the noises in this publications are
practically the same. The only difference is in dimensionless
multiplier which depends on the ratio of the substrate (the
mirror) and coating material constants. But substantial difference
in \cite{BV} and \cite{Row} is in the numerical estimate of the
spectral densities of the noise. This difference appears because
the numerical value of the thermal expansion factor of tantalum
pentoxide $\alpha_{\rm Ta_2\, O_5}$ in \cite{BV} has been borrowed
from the experiment by Inci \cite{Inchi}: $\alpha_{\rm Ta_2\,
O_5}=[-4.43\pm0.05]\times 10^{-5} \rm\:K^{-1}$. This value is
unexpectedly negative and large in contrast to \cite{Row}, where
the value of $\alpha_{\rm Ta_2\, O_5}=+3.6\times 10^{-6}
\rm\:K^{-1}$ has been borrowed from the measurement by Tien et al.
\cite{charly} (positive value but without confidence limits). Due
to this difference the predictions of the value of the noise
produced by thermoelastic effect in \cite{BV} and \cite{Row}
differs by one order. It is worth noting that this difference
between these independent measurements may not be due to a mistake
(or mistakes) of experiment but due to ''pathological'' properties
of thin tantalum pentoxide layers. This harsh description of that
solid belongs to S.~Witcomb \cite{wit}, who mentioned that the
properties of $\rm Ta_2\, O_5$ layer substantially depend on the
chosen procedure of the layer deposition, and one may expect even
mechanical and optical anisotropy of this type of layer.

These circumstances are the motivation for the experiments we have
carried out and which are described below.

\section{The procedures and results of measurements.}

To measure the value of thermal expansion factor of tantalum
pentoxide $\alpha_{\rm Ta_2 \,O_5}$ we used a well known method
(see e.g. \cite{charly, tow}). This method is based on the
measurement of bending of thin substrate plate (made of $\rm Si \,
O_2$ in our case) with coating, when plate is heated. The bending
appears due to the difference $\alpha_{\rm Ta_2 \,O_5}-\alpha_{\rm
Si \,O_2}$. If one end of the plate, which length is $l$, is
rigidly attached to massive platform (which is not distorted by
change of temperature) then one can observe the displacement
$\Delta z$ of the plate free end due to the tension between the
coating and the substrate that appears if the plate temperature
changes by $\Delta T$. The value of $\alpha_{\rm Ta_2 \,O_5}$ may
be calculated using the following formula which can be easily
derived from formulas (30c) and (32b) of \cite{tow}:
\begin{equation}
\alpha_{\rm Si \,O_2}-\alpha_{\rm Ta_2 \,O_5}=\frac{1}{3}
\frac{E_{\rm Si \,O_2}(1-\nu_{\rm Ta_2 \,O_5})d_{\rm Si
\,O_2}^2}{E_{\rm Ta_2 \,O_5}(1-\nu_{\rm Si \,O_2}) d_{\rm Ta_2
\,O_5} l^2}\left[\frac{\Delta z}{\Delta T}\right] \label{fin}
\end{equation}
where $E_{\rm Si \,O_2}$ is the plate Young modulus, $E_{\rm Ta_2
\,O_5}$ is the $\rm Ta_2 \,O_5$ Young modulus, $d_{\rm Ta_2
\,O_5}$ is the total thickness of $\rm Ta_2 \,O_5$ coating,
$d_{\rm Si \,O_2}$ is the thickness of the plate, $\nu_{\rm Ta_2
\,O_5}$ and $\nu_{\rm Si \,O_2}$ are the Poisson's ratios of $\rm
Ta_2 \,O_5$ and plate solid respectively.

To reproduce the coating for a real mirror manufactured for the
LIGO antenna in our tests we have prepared $100 \rm\: \mu m$ flat
fused silica plates which were very well polished and have the
shape of stripes $2\rm\: cm$ long and $1\rm\: cm$ wide. A set of
such thin fused silica plates-stripes was placed on a big
relatively thick fused silica mini platform. The very ends of each
stripe were rigidly attached to the platform. Due to the care of
Dr. H.~Armandula this platform (with the plates-stripes on it) was
placed into the same vacuum chamber together with big End Mirror
for LIGO ($10\rm\:cm$ thick, $25\rm\:cm$ in diameter), where the
deposition of the multilayer coating was performed. Our fused
silica plates-stripes ''received'' the same 19 pairs of
quarter-wavelength $\rm Ta_2\,O_5+Si\,O_2$ layers. Thus our
plates-stripes were covered with the same coating (which has the
thickness of $5.5\rm\:\mu m$ and the total thickness of $\rm
Ta_2\,O_5$ -- $2.2\rm\:\mu m$) as the big mirror. This coating
provides the value of $(1-R)\simeq 1\times 10^{-5}$.

When the process of deposition was over and the plates-stripes
were detached from the platform we have got the first (unexpected)
experimental result: all plates-stripes were bent. The measured
value of the curvature radius of these plates-stripes was $\simeq
11 \rm\: cm$ (in the plane of long axis of the stripes). This
bending remained unchanged many weeks after the detachment from
the platform.

The main elements of the measuring set up are the two massive
fused silica rectangular blocks which were rigidly attached to a
heavy optical bench. A plate-stripe was rigidly clamped between
two blocks. To evade damage due to the clamping a $15\rm\: \mu m$
teflon film was inserted between plate-stripe and the surfaces of
fused silica blocks. Power stabilized He-Ne laser, two lenses and
photodiode were installed on the bench. This simple optical system
permitted to measure $0.1 \rm\: \mu m$ displacement of the
plate-stripe edge (''knife and slot'' technique). Small box around
a plate-stripe permitted to heat the plate-stripe by a slow stream
of heated air.

Four sets of measurements with two different plates were carried
out. The measurements with each plate were performed two times,
second time` with the plate flipped up side down. The value of
$[\Delta z/\Delta T]$ was obtained from the directly measured
dependence of the edge displacement at the free end of
plate-stripe on temperature. By using formula (\ref{fin}) with
parameters $\alpha_{\rm Si\, O_2}=5.5\times10^{-7}\rm\: K^{-1}$,
$E_{\rm Si\, O_2}=7.2\times10^{11}\rm\: \frac{erg}{cm^3}$,
$\nu_{\rm Si\, O_2}=0.17$, $E_{\rm Ta_2\,
O_5}=1.4\times10^{12}\rm\: \frac{erg}{cm^3}$, $\nu_{\rm Ta_2\,
O_5}=0.23$ and measured $[\Delta z/\Delta T]$ the linear thermal
expansion coefficient $\alpha_{\rm Ta_2\,O_5}$ was estimated. Its
value is equal to:
$$\alpha_{\rm Ta_2\, O_5}=[5\pm 1]\times10^{-6}\rm\: K^{-1}.$$

\section{Conclusion.}

The obtained value of $\alpha_{\rm Ta_2\, O_5}$ is evidently
closer to the one obtained by Tien et al. \cite{charly} then to
the one obtained by Inci \cite{Inchi}. In the same time our value
is almost two times larger then the smaller value \cite{charly}.
In addition we have to note that our value of $\alpha_{\rm Ta_2\,
O_5}$ is based not only on the measured value of $\Delta z$ but
also on assumption that the values of $E_{\rm Si\, O_2}$ and
$E_{\rm Ta_2 \, O_5}$ (measured by other experimentalists) are
correct. Taking into account that these values may be
substantially different in thin layers and that the process of
deposition may also cause changes of the Young moduli values (see
e. g. \cite{Martin}), we may conclude that the value of
$\alpha_{\rm Ta_2 \, O_5}$ is between $3\times 10^{-6}\rm\:K^{-1}$
and $7\times 10^{-6}\rm\:K^{-1}$. The only ''remedy'' to solve
this problem is the direct measurement of the mirror with coating
surface fluctuations spectral density. At the same time the limits
(from $3\times 10^{-6}\rm\:K^{-1}$ to $7\times
10^{-6}\rm\:K^{-1}$) permit to cancel the pessimistic prediction
for the spectral density of the thermoelastic noise in the coating
\cite{BV} which was based on Inci's value $\alpha_{\rm Ta_2\,
O_5}=[-4.43\pm0.05]\times 10^{-5} \rm\:K^{-1}$. For example, the
potentially achievable sensitivity for LIGO-II near the frequency
of observation $f\simeq10^2 \rm\: Hz$ may not be at the level of
$S_h\simeq 1.5 \times 10^{-23} \rm\:1/\sqrt{Hz}$ (as in \cite{BV})
but between $0.6 \times 10^{-24} \rm\:1/\sqrt{Hz} \leqslant S_h
\leqslant 1.4 \times 10^{-24} \rm\:1/\sqrt{Hz}$. It means that the
projected sensitivity of LIGO-III (better than the Standard
Quantum Limit) will be evidently not possible due to this noise.

To the above conclusive remarks we have to add that the
''unpleasant discovery'' i.e. the observation of the ''frozen''
bending of the plates-stripes with coating, which indicates the
existence of mechanical stresses in the coating has to be regarded
as strong argument in favor of direct measurement of the noises in
coating, especially the measurement of $1/f$ component.

The authors of this article are expressing their sincere gratitude
to S.P.~Vyatchanin and H.~Armandula for their advice and direct
help. This work was supported by NSF grant PHY-0098-715 and by
Russian Ministry of Industry and Science.

\end{document}